# AlN Combined Overtone Resonators for the 5G mmWave Spectrum

Guofeng Chen and Matteo Rinaldi
Northeastern University,
Boston, USA

*Abstract*—This work presents a new acoustic MEMS resonator technology, dubbed Aluminum Nitride (AlN) Combined Overtone Resonator (COR), capable of addressing the filter requirements for the 5G mmWave spectrum in the 6-40GHz range. The COR exploits the multimodal excitation of two higher-order Lamb waves (2nd and 3rd order Asymmetrical Lamb Waves) in a suspended thin-film AlN plate to transduce a 2-dimensional vibration mode with high electromechanical coupling coefficient $k_t^2$ (up to 1.9%) and quality factor $Q$ >1100 at twice the frequency of a fundamental thickness-extensional mode in the same structure. Analytical and finite-element method (FEM) models are developed to describe the working principle of the COR technology and predict the achievable $k_t^2$, $Q$ and lithographic frequency tunability. An 8.8 GHz COR prototype was fabricated showing a high $k_t^2$~0.3% (using a simple top-electrode-only configuration with a 2-mask process) and a groundbreaking $Q$~1100 which is the highest ever achieved among piezoelectric resonators above 6 GHz. The $f$-$Q$ product ~1×10¹³ is the highest among all demonstrated piezoelectric resonators with metallic coverage >50%. Additionally, the capability of the COR technology to deliver contiguous filters with bandwidths between 355 and 592 MHz (aggregated BW>2GHz) in the mmWave spectrum, with relaxed lithographic requirements, is demonstrated by FEM.

*Index Terms*—5G mobile communication, electromechanical coupling coefficient, millimeter wave communication, overtone, piezoelectric resonator filter, quality factor

## I. INTRODUCTION

The development of wireless communication systems has been steadily growing and conventional sub-6 GHz frequency bands are too congested to meet the high data rate requirements of several emerging technologies. For example, connected-vehicle-to-everything communication (V2X) systems have been recently developed to allow vehicles to communicate with moving parts of the traffic system around them to increase awareness of all surroundings, reduce the risk of collisions, and maximize the transportation efficiency. This game-changing technology requires high-bandwidth, low-latency and high-reliability short-range wireless links to communicate with compatible systems on vehicles, pedestrians and infrastructures. The 5G cellular network can be considered the key enabler for such a ubiquitous and pervasive mobile internet connectivity. In particular, the use of the millimeter wave (mmWave) spectrum represents the major leap forward in the 5G network as it enables improvements in data speed, capacity, quality and latency that are unimaginable in 3G and 4G networks. 5G frequency bands from 6 GHz to 40 GHz are typically referred to as cm-mmWave in the industry. Unlike sub-6 GHz spectrum with BW typically between 5 MHz and 20 MHz, 5G cm-mmWave spectrum provides contiguous bandwidths (BWs) from 40 MHz up to 2 GHz. This high-band spectrum enables data rates in the tens of Gbps range with extremely low latency, providing significant opportunities for very high throughput services such as, Enhanced Mobile Broadband (eMBB), Ultra Reliable Low Latency Communications (URLLC) and Massive Machine Type Communications (mMTC) [1]. It is worth noting that the first set of high-band auction, concluded by the Federal Communications Commission (FCC) in May 2019, offered more than 2,900 Upper Microwave Flexible Use Service licenses in the 24 GHz band: the lower segment of the 24 GHz band (24.25–24.45 GHz) is licensed as two 100-megahertz blocks, while the upper segment (24.75–25.25 GHz) is licensed as five 100-megahertz blocks. These 5G 24 GHz bands are closely located to the 23.8 GHz band which is used for sensitive meteorological and oceanographic measurements, therefore, the adoption of these 5G bands in communication systems requires the use of pass-band filters with a relatively small fractional bandwidth of ~0.42% (i.e. 100 MHz) and a high $Q$ > 500 to achieve the steep roll-off and the large out-of-band rejection needed to enable coexistence with the adjacent band.

Therefore, new solutions are highly needed to address this incoming 5G mmWave filter challenge. The capability of synthesizing high-performance filters for the mmWave would be groundbreaking as it would enable the miniaturization and integration of 5G systems for a truly ubiquitous and pervasive mobile internet connectivity.

Although the electromagnetic (EM) wavelengths are significantly reduced at mmWave, the dimensions of conventional EM based filters are still more than 600 times larger than what could be achieved with the acoustic counterparts, making them not suitable for the implementation of next generation miniaturized mobile devices [2]–[4]. Besides, the low quality factor of EM filters results in poor roll-off which renders the system vulnerable to crosstalk from adjacent channels.

Micro-electro-mechanical system (MEMS) components, especially Surface Acoustic Wave (SAW) and Bulk Acoustic Wave (BAW) resonators, have been explored and widely employed as radio frequency (RF) filters in current mobile devices for frequency band selection due to the high-quality factor, high electromechanical coupling coefficient (corresponding to high fractional BW) and small form-factor they can achieve in the sub-6 GHz range. Nevertheless, when scaled above 6 GHz, all the existing micro-acoustic resonator technologies resonators have suffered critical limitations associated with increased acoustic losses and aggressively scaled dimensions that have prevented the synthesis of high performance filters based on these technologies [5].

SAW filters operating at 15 GHz were demonstrated employing 145 nm wide interdigital (IDT) electrodes. However, an insertion loss >40 dB was measured for these devices due to their heavily degraded $Q$ (commercially available SAW filters show I.L. < 1 dB in 1 GHz range) [6]. Aluminum nitride (AlN) BAW resonators in X band and K band were also demonstrated. Although these high frequency BAW resonators maintain an electromechanical coupling coefficient ($k_t^2$) > 6% (enabled by the high piezoelectric coefficient, $e_{33}$, employed to transduce vibration), they show very low $Q$-values (~300) which are 10 times lower than the ones typically achieved by the same technology in the sub-6 GHz frequency bands, resulting in the synthesis of filters with high insertion loss of 3.8-11 dB (I.L.$\propto k_t^2 \cdot Q$) [2], [7], [8]. This drastic performance degradation is associated with the fact that ultra-thin piezoelectric and metal layers are required to achieve thickness-extensional vibration at mmWave frequencies. For example, the excitation of a 24 GHz thickness-extensional mode of vibration requires the use of an ultra-thin material stack composed of a 120 nm thick AlN piezoelectric layer and 17 nm thick ruthenium (Ru) electrodes. Studies have shown that the first few tens of nanometers of a deposited AlN thin-film are not properly oriented along the c-axis, resulting in degraded piezoelectric coupling and quality factor $Q$ [9]–[11]. In particular, it was experimentally demonstrated that the X-Ray diffraction (XRD) full width at half-maximum (FWHM, an indication of crystal quality with smaller values corresponding to better crystal orientation) increases dramatically when the deposited AlN film is thinner than 250 nm. For example, while the FWHM increases by 10% when the AlN film is scaled from 1000 nm to 250 nm, the degradation is more than 20% when it is scaled to 100 nm. For these reasons, it is certainly preferable to employ AlN films thicker than 200 nm in the make of micro-acoustic resonators in order to avoid the consequences of thin-film material degradation. In addition to the issues associated with the use of reduced quality ultra-thin piezoelectric films, the performance of the resonator is also significantly affected by the use of ultra-thin metal electrodes that unavoidably introduce a large electrical resistance that dramatically reduces the loaded $Q$ of the resonator [12]. For these reasons, to date AlN BAW resonators, or any other piezoelectric resonators, have not been successfully scaled to operate above 6 GHz without a significant performance degradation which has prevented the implementation of performing mmWave filters.

In order to synthesize performing mmWave filters, resonators with high $Q$-values larger than 500 are required. The use of high $Q$ resonators enables the achievement of the steep roll-off required for the implementation of contiguous filters separated by a minimum guard band for efficient spectrum usage. The use of high-$Q$ resonators also enables more flexibility in the filter design with trade-off possibilities between filter rejection and bandwidth. Furthermore, the insertion loss (I.L.) heavily depends on the resonator $Q$ when resonators with relatively modest $k_t^2$ values are employed to synthesize relatively narrow fractional bandwidth filters (FBW) ($k_t^2$<0.8 % for 100 MHz BW at 24 GHz) [13].

In this context, overtone resonators are intrinsically advantageous as they typically achieve high $Q$ at very high frequencies due to reduced air damping and acoustic loss [14]–[18]. However, their $k_t^2$-values are dramatically reduced as they scale inversely proportional to the cube of the order number.

In this work, a new class of acoustic resonators, Aluminum nitride (AlN) *Combined Overtone Resonators* (CORs), is presented to overcome the aforementioned fundamental challenges. CORs relies on the piezoelectric multimodal excitation of two higher order Lamb waves (the $2^{nd}$ and $3^{rd}$ order Asymmetrical Lamb Waves, i.e. A2 and A3) to transduce a 2-dimensional (2D) mechanical mode of vibration [19], [20] in the cross-section of a suspended thin-film AlN plate. For the first time, a combination of multiple overtone modes is employed to achieve a more efficient piezoelectric transduction. Exploiting the multimodal excitation of combined overtones, CORs can operate in the mmWave spectrum while maintaining relaxed lithographic requirements (minimum feature > 100 nm) and relatively thick AlN films (>220 nm), which eliminate the performance degradation issues associated with ultra-thin film materials and directly translates in the achievement of high $Q$ > 1000. Furthermore, thanks to the coherent combination of the $e_{33}$ and $e_{15}$ piezoelectric coefficients of AlN employed to transduce the combined overtone 2D mechanical mode of vibration, CORs achieve relatively high electromechanical coupling coefficient $k_t^2$ (0.8% to 1.9%) despite the use of higher order modes (i.e. overtones) in the structure. It is also worth noting that due to the dependence of this combined overtone 2D vibrational mode on both the vertical and the lateral dimensions of the structure, the resonant frequencies of CORs can be lithographically tuned to synthesize monolithic multi-frequency filters on the same chip with minimal fabrication complexity. Therefore, differently from any existing resonator technology, mmWave CORs can simultaneously achieve high $Q$-values >1000 and relatively high electromechanical coupling coefficient $k_t^2$ (0.8%~1.9%) suitable for the implementation of contiguous filters (bandwidth from 300 to 592 MHz) for aggregated bandwidth from 1.6 to 2.6 GHz on the same substrate in the 24-40 GHz frequency range.

## II. PRINCIPLE OF OPERATION

Lamb wave theory has been studied to identify all vibration modes that can be described by a one-dimensional (1-D) displacement vector in a plate [20]. This theoretical model has been an important tool for analyzing and optimizing Lamb wave resonators and it is also adopted here to describe the 2-D motion of Aluminum nitride (AlN) Combined Overtone

Resonators (CORs).

AlN CORs are formed by a suspended AlN thin film and rely on the use of one (i.e. top electrode-only configuration) or two (i.e. top and bottom electrodes configuration) interdigital transducer (IDT) electrodes for the piezoelectric multimodal excitation of the 2$^{nd}$ and 3$^{rd}$ order asymmetrical Lamb-wave overtones (A2 and A3 respectively) resulting into a 2-dimensional (2D) combined overtone mode (COM) of vibration in the cross-section of a suspended thin-film AlN plate. This COM is excited in the structure when the thickness of the AlN layer, $t_{AlN}$, is approximately equal to the pitch, $W$, of the employed IDT. Such a COM is characterized by a 2D displacement vector with components along both the lateral ($\widetilde{\mu_x}$) and vertical ($\widetilde{\mu_z}$) directions of the plate. It is worth noting that $\widetilde{\mu_x}$ corresponds to the vibration modeshape of the A3 mode while $\widetilde{\mu_z}$ corresponds to the vibration modeshape of the A2 mode. The general expression of ($\widetilde{\mu_x}$) and ($\widetilde{\mu_z}$) for the A2-A3 COM is reported in (1), assuming infinite periodic boundary conditions in both directions.

$$\begin{bmatrix} \widetilde{\mu_x} \\ \widetilde{\mu_z} \end{bmatrix} = \begin{bmatrix} \bar{X}A(x)B(z) \\ \bar{Z}C(x)D(z) \end{bmatrix} = \begin{bmatrix} -\bar{X}cos(\beta_x x)\,cos(\beta_{z\_x}z) \\ -\bar{Z}\,sin(\beta_x x)\,cos(\beta_{z\_z}z) \end{bmatrix} \quad (1)$$

Where $\bar{X}$ and $\bar{Z}$ are the magnitudes of the displacement components along the $x$- and $z$- directions, respectively; $\beta_x$ is the wave-vector relative to the motion along the $x$-directions for both $\widetilde{\mu_x}$ and $\widetilde{\mu_z}$; $\beta_{z\_x}$ and $\beta_{z\_z}$ are the wave-vectors relative to the motion along the $z$-directions for $\widetilde{\mu_x}$ and $\widetilde{\mu_z}$, respectively. The expressions describing the three wave-vectors are reported in (2), (3), and (4).

$$\widetilde{\beta_x} = \frac{\pi}{W} \quad (2)$$

$$\widetilde{\beta_{z\_x}} = \frac{3\pi}{t_{AlN}} \quad (3)$$

$$\widetilde{\beta_{z\_z}} = \frac{2\pi}{t_{AlN}} \quad (4)$$

where $W$ is the pitch of interdigital transducer (IDT) electrodes employed to transduce the COM and $t_{AlN}$ is the thickness of the AlN plate. It is worth noting that all the three wave vectors have different mode orders, resulting in vortexes in the total displacements field.

The vibration modeshape of the COM was simulated by finite-element method (FEM) in COMSOL, as shown in Fig. 1. The simulation was performed using two periodic unit cells (finger number $N$=2) with periodic boundary conditions along $\hat{x}$ and stress-free boundary conditions in $\hat{z}$. The FEM simulated modeshape of the COM in Fig. 1 matches well to the analytical description by equations 1-4 with small deviations along the z-direction due to the lack of periodic boundary conditions (i.e. stress-free boundaries on the top and bottom surfaces of the FEM model). The FEM simulated modeshapes of both $\widetilde{\mu_x}$ and $\widetilde{\mu_z}$ are also reported in Fig. 1 highlighting the clear matching with the displacement profiles of the A3 and A2 modes, respectively.

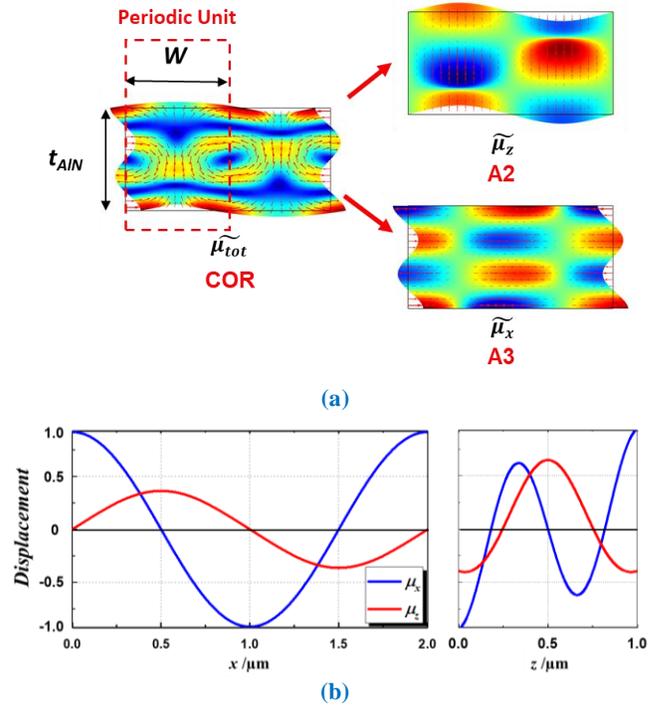

Fig. 1. a) The FEM-simulated modeshape of two periodic units ($N$=2) of an AlN Combined Overtone Resonator (COR) for $t_{AlN}$=1 μm and $W$ = 1 μm. This COR operates at 9771 MHz. b) The displacement components in x- and z-directions. The effect of electrodes is not included in the simulations. Normalized $\widetilde{\mu_x}$ and $\widetilde{\mu_z}$ along the lateral direction (left) and the thickness direction (right) of COR.

## A. Piezoelectric Coupling Efficiency

Multiple parameters have been used in the literature to describe the piezoelectric coupling efficiency of resonators [21]. Particularly important is the electromechanical coupling coefficient, $k_t^2$, which is a quantitative measure of the conversion between the electrical and mechanical energy in the electromechanical resonator:

$$k_t^2 = \frac{\pi^2}{8} \frac{f_p^2 - f_s^2}{f_s^2} \qquad (5)$$

where $f_s$ and $f_p$ are the series and parallel resonant frequencies of the resonator, respectively.

Another important parameter is the piezoelectric coupling constant, $K^2$, which identifies the maximum $k_t^2$ that can be achieved for any mode of vibration through optimal excitation [22], [23]. $K^2$ can be evaluated using (5) with $f_s$ and $f_p$ replaced by $f_o$ and $f_m$, as in (6), where $f_o$ and $f_m$ are the non-metallized and the metallized resonant frequencies of the resonator, respectively. Therefore, $K^2$ can be readily computed using the dispersion characteristics of the resonator [24], [25], [26].

$$K^2 = \frac{\pi^2}{8} \frac{f_o^2 - f_m^2}{f_m^2} \qquad (6)$$

In this work, FEM simulations were performed in COMSOL Multiphysics to study the piezoelectric coupling constant of the COM in a suspended AlN plate. Both the resonant frequencies and the $K^2$ values were computed for a 1 µm thick AlN while varying its lateral dimension $W$ (i.e. the wave number $k_x$) (Fig. 2). As evident, a maximum $K^2$-value of ~1.2% is achieved when the AlN thickness, $t_{AlN}$, is approximately equal to $W$. It is worth noting this $K^2$-value is the highest ever predicted for any overtone mode excited in an AlN plate. It is worth noting that the presence of the metal electrode, which is not included in this analysis, but it is required in an actual COR implementation, further increases the maximum achievable $k_t^2$ as explained in the following sections.

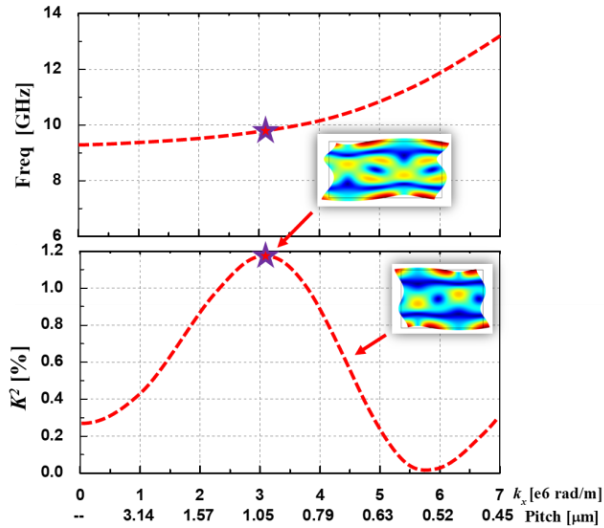

Fig. 2. Dispersion curve and $K^2$-values of the COM excited in a 1 µm-thick AlN plate for different values of its width ($W$). The highest $K^2$ value and its corresponding frequency are marked by the stars. The modeshapes for $W=1$ µm and $W=0.7$ µm are shown in the insets.

Generally, AlN resonators rely on one or more piezoelectric coefficients such as, $e_{31}$, $e_{33}$, and $e_{15}$, to transduce a mechanical mode of vibration [27]–[30]. Since the COM is characterized by a 2D displacement vector with a shear vibration component along the lateral direction ($\widetilde{\mu_x}$) and a longitudinal vibration component along the vertical ($\widetilde{\mu_z}$) direction of the plate, the coherent combination of the $e_{15}$ and $e_{33}$ piezoelectric coefficients of AlN is exploited for its transduction.

The analytical model proposed by Berlincourt et. al. [31], [32] was considered to investigate the effect of different excitation schemes on the $k_t^2$-values of a device with a volume $\Omega$:

$$k_t^2 = \frac{\pi^2}{8} \frac{U_{coupling}^2}{U_m U_e} \qquad (7)$$

where

$$U_m = \frac{1}{2} \int_\Omega T_i s_{ij}^E T_j d\Omega \qquad (8)$$

$$U_{coupling} = \frac{1}{4} \int_\Omega (T_i e_{ni} E_n + E_n e_{ni} T_i) d\Omega \qquad (9)$$

$$U_e = \frac{1}{2} \int_\Omega E_m \varepsilon_{mn}^T E_n d\Omega \qquad (10)$$

Here $U_m$ is the mechanical energy, $U_{coupling}$ is the energy associated with the coupled electrical mechanical domain (mutual energy) and $U_e$ is the electrical energy carried over the resonator volume $\Omega$. $m$, $n$=1, 2, 3 and $i$, $j$=1, 2, 3, 4, 5, 6.

$E_m$ is the component of the electric field vector. $T_i$ is the component of the stress. $e_{mn}$, $\varepsilon_{mn}^T$ and $s^E{}_{ij}$ are the piezoelectric coefficient, the dielectric constant and the elastic compliance, respectively.

Equations (7)-(10) provide important insights for the design and the optimization of the excitation scheme employed to transduce vibration in the structure. In particular, (9) indicates that the higher is the integral of the product of electric and stress fields in space the higher is the $k_t^2$. Therefore, the design of an excitation scheme providing an electric field distribution matching the stress field of the COM in the structure is critical to maximize the coupling.

Similarly to more conventional BAW and Lamb-wave resonators, CORs can use two main kinds of excitation schemes: the Thickness-Field-Excited (TFE) configuration, using both top and bottom IDT electrode, and the Lateral-Field-Excited (LFE) configuration, using only a top IDT (or only a bottom IDT) electrode [33]. The LFE configuration employs metal fingers with alternating electrical polarities on the top (or the bottom) of the resonator, as shown in Fig. 3(a). The typical TFE configuration (referred as TFE-1) employs metal fingers with inversed electrical polarity between top and bottom, as shown in Fig. 3(b). An unconventional TFE configuration (referred as TFE-2) is also proposed and examined here which employs metal fingers with the same electrical polarities on the top and the bottom of the resonator, as shown in Fig. 3(c).

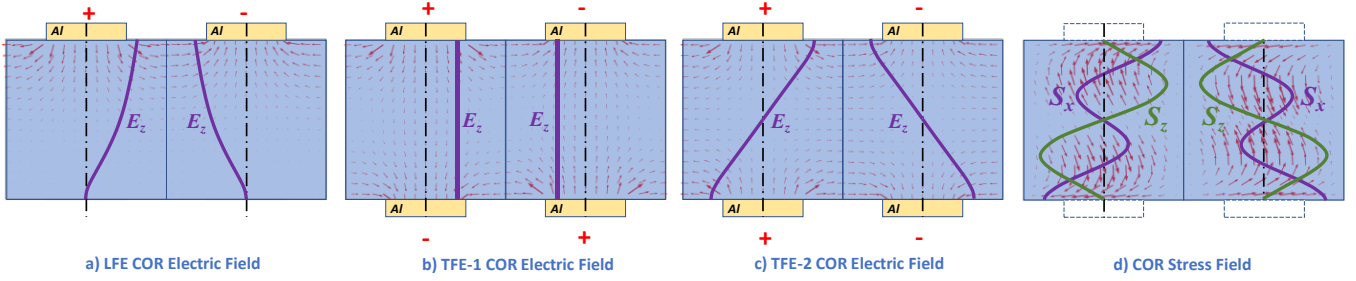

Fig. 3. Electric field distributions for electrical configurations (a) LFE, (b) TEF-1 and (c)TFE-2. The directions and length of red arrows repent the direction and strength of the electric field vectors. The bold purple lines represent the electric field component $E_z$ (in the mid-plane of each unit) across the thickness, and on the right side of the dash-dotted line means positive sign and vice versa. The asymmetric stress field distributions for CORs without the mass loading of the IDT is also shown in (d).

The choice of the excitation scheme affects the electric field distributions in the piezoelectric film and therefore the value of $k_t^2$ for a given mode of vibration (i.e. given stress field distribution). The electric field distributions for the 3 different excitation schemes were simulated by FEM and reported in Fig. 3 (a-c) while the simulated stress field distribution of a COM, neglecting the IDT mass loading effect, is plotted in Fig. 3 (d). The simulation results indicate that the conventional symmetric electric field distribution provided by the TFE-1 scheme is not suitable for the transduction of the COM as it results in a mutual energy, $U_{coupling}{\sim}0$ [34]. On the other hand, the asymmetrical electric field distribution provided by the LFE and TFE-2 configurations match the asymmetrical stress field of the COM resulting in a non-zero mutual energy value. In particular, the proposed unconventional TFE-2 excitation scheme best matches the stress field of the COM enabling, for the first time, the efficient piezoelectric excitation of even-order overtones (i.e. A2, A4 etc.) which results in a high electromechanical coupling coefficient $k_t^2$ for the COM.

The effect of the electrode mass loading on the stress field distribution was also analyzed by FEM. The simulated electric field component, $E_z$, and stress field components, $S_x$ and $S_z$, in the center of a periodic unit cell are plotted in Fig. 4. The simulation results clearly show that the mechanical loading of the metal electrodes modifies the stress field distribution of the COM improving further the matching with the electric field distribution provided by the TFE-2 scheme (particularly in the middle region of the material stack). As a result, significantly higher $k_t^2$-values can be achieved in an actual COR implementation employing metal IDT electrodes.

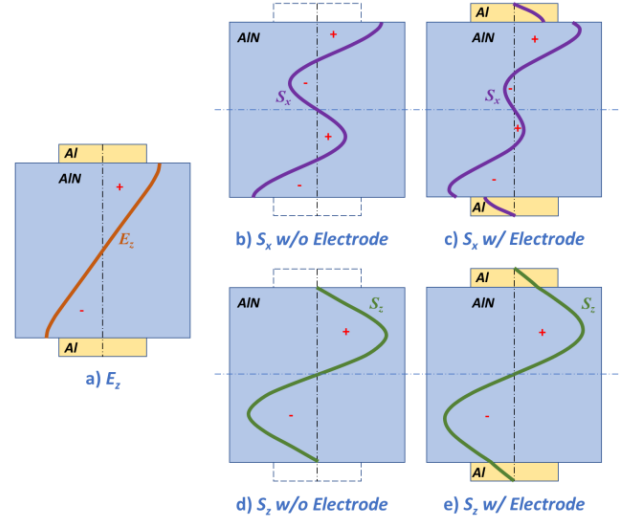

Fig. 4. (a) $E_z$ for TFE-2; $S_x$ distribution in the center of a periodic unit of CORs (b) without and (c) with the mechanical effects of electrodes; $S_z$ distribution in the center of a periodic unit of CORs (d) without and (e) with the mechanical effects of electrodes;

The choices of the electrode materials and thicknesses are also important factors affecting the device $k_t^2$. Ruthenium (Ru), molybdenum (Mo), and tungsten (W) are widely adopted electrode materials for BAW resonators as they provide good crystal interfaces for the growth of highly c-oriented AlN with excellent piezoelectric properties. However, the use of these metals significantly decreases the device resonant frequency due to their high density and stiffness. It was recently reported that high quality AlN thin-films (FWHM=1.05° [35]), as good as the one obtained on top of Ru, Mo or W, can be attained also on top of lighter metals, such as aluminum (Al), representing a valid alternative for the implementation of resonators operating above 6 GHz. For this reason, Al was chosen as electrode material for the analysis and the experiments presented in this work.

Different thicknesses of the Al electrodes were simulated to find the optimal value that maximizes the $k_t^2$-values for the TFE-2 configuration. As shown in Fig. 5(a), a maximum $k_t^2$-value of 1.9% was found for an optimal ratio of AlN to Al of 1000 nm to 110 nm for the TFE-2 CORs. This represents a more than 60% increase in $k_t^2$ compared to the electrode-less case (Fig. 2). As a comparison, resonators relying on fundamental modes of vibrations (such as conventional BAW resonators) typically show a 14% increment in $k_t^2$ associated to the presence of metal electrodes [31]. A similar study was performed for the LFE configuration (top Al IDT electrode only) showing a maximum $k_t^2$-value of 0.8% for optimized vertical and lateral dimensions (IDT pitch $W$ and metallic coverage $\alpha$) of the resonator stack, as shown in Fig. 5(b).

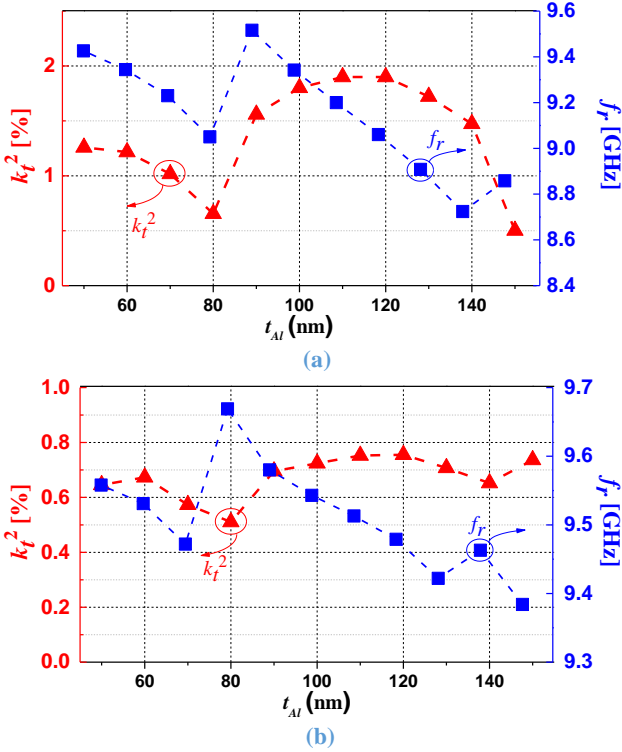

Fig. 5. $k_t^2$-values and resonant frequencies $f_r$ for different Al thickness ($t_{Al}$) of CORs based on (a) TFE-2 and (b) LFE configurations. AlN thickness ($t_{AlN}$) is fixed at 1 μm. $W$=1.1 μm, $N$=2, $\alpha$=50%, $t_{Al}$ =110 nm for TFE-2. $W$=1.15 μm, $N$=2, $\alpha$=35%, $t_{Al}$ =120 nm for LFE.

### B. Quality Factor Q

The quality factor $Q$ of a piezoelectric resonators is defined as the ratio between stored and dissipated energies. The total quality factor $Q_{total}$ can be generally expressed as in (11) [36], [37],

$$\frac{1}{Q_{total}} = \frac{1}{Q_{anchor}} + \frac{1}{Q_{interface}} + \frac{1}{Q_{material}} + \frac{1}{Q_{electrical}} + \frac{1}{Q_{dielectric}} + \frac{1}{Q_{intrinsic}}$$ (11)

The anchor loss, $1/Q_{anchor}$, describes the acoustic energy lost through the anchor to the substrate. The typical frame designs for energy confinement employed for lower frequency resonators cannot be readily applied to higher frequency devices due to their reduced dimensions [31], [38]. As can be seen from the modeshape in Fig. 1, COM has a strong shear vibration which is characterized by high $Q$ [39], [40]. Therefore, the energy is naturally well confined inside the resonator body [41]. The interface loss, $1/Q_{interface}$, is caused by the plane stress jump at the interface between the piezoelectric layer and the metal electrode. As shown in the COR modeshape simulation in Fig. 1, the IDT electrodes are placed in the center of each periodic unit where the displacement is minimum, which minimizes the scattering of the acoustic wave at the electrode-AlN interface and the associated damping. The material loss, $1/Q_{material}$, is related to the presence of defects in the crystal. Since high operating frequency is achieved in CORs while using an AlN layer thicker than the one employed by more conventional technologies operating in the same frequency range, the material loss is mitigated in CORs. The electrical loss, $1/Q_{electrical}$, is determined by the ohmic losses of the metal electrodes and the electrical routing. The use of overtones in a relatively thick AlN layer allows the use of sufficiently thick metal electrodes with minimal electrical loading effect. The dielectric loss, $1/Q_{dielectric}$, is determined by the inherent dissipations of electromagnetic energy in the dielectric material which are reduced in the sufficiently thick AlN films employed in CORs. In fact, it was experimentally shown that the dielectric loss increases rapidly only when the AlN film is scaled below 250 nm [23]. Other dissipative mechanisms intrinsic to the resonating material, $1/Q_{intrinsic}$, consist of thermoelastic, phonon-electron, and phonon-phonon interactions [42], [43]. The Akheiser theory [43] predicts that the $f$-$Q_{intrinsic}$ product of AlN increases linearly with frequency for frequencies larger than ~1 GHz, enabling the achievement of high $Q_{intrinsic}$ in cm-mmWave AlN resonators.

In summary, it can be qualitatively concluded that most of the energy loss mechanisms are mitigated in CORs thanks to the use of overtones in a relatively thick material stack.

### III. FABRICATION AND EXPERIMENTAL RESULTS

As a technological proof-of-concept LFE CORs were designed, fabricated and tested. Although the LFE configuration yields a lower $k_t^2$ than the TFE-2 one, it is characterized by the minimal fabrication complexity. The devices were fabricated using a simple 2-mask fabrication process as described in Fig. 6. The process started with the sputter deposition of a highly textured (1.1° FWHM) 1 μm thick AlN film directly on top of a polished high-resistivity Silicon wafer. Then an Aluminum (Al) layer was sputter deposited and patterned (using an i-line stepper) on top of the AlN to form the device top IDT electrode with a pitch of 1.15 μm. The AlN film was then dry etched in Inductively Coupled Plasma (ICP) to define the shape of the suspended AlN plate. Finally, the devices were released from the substrate by XeF$_2$ dry isotropic etching of silicon. A Scanned Electron Microscope (SEM) picture of a fabricated COR is shown in Fig. 7. As evident in Fig. 7(c), a sloped sidewall (SW) angle around 60° was observed in the cross-sectional view due to the photoresist etching mask employed in the process.

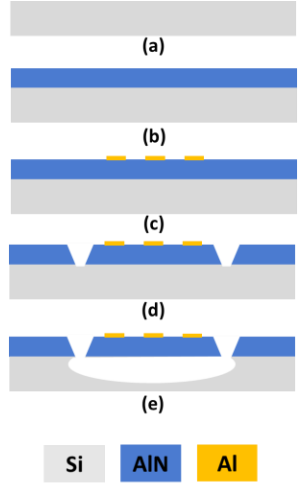

Fig. 6. Microfabrication process for LFE CORs.

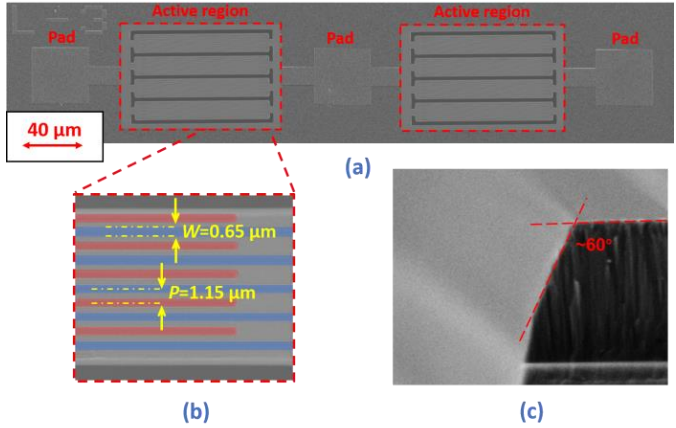

Fig. 7. SEM image of (a) CORs, (b) zoom-in of COR IDT, and (c) etching sidewall of COR. The device consists of an array of 8 resonators to obtain a high static capacitance value for matching 50 Ω termination.

The performance of the fabricated prototypes were characterized using a Microwave Network Analyzer (Keysight N5221A PNA) in air and at room temperature. The measured admittance is shown in Fig. 8(a). A strong resonance, corresponding to the transduction of the combined overtone mode of vibration, was measured near the frequency predicted by the FEM simulations with an error of 3%. The measured COR showed a loaded quality factor $Q_l$ of ~750 and a mechanical quality factor $Q_m$ of ~1100 (corresponding to $f \cdot Q_m \sim 1 \times 10^{13}$). Such $Q$-values are the highest ever reported for piezoelectric micro-acoustic resonators operating > 6 GHz. In addition, the measured $f \cdot Q_m$ product ~$1 \times 10^{13}$ is the highest ever demonstrated for piezoelectric micro-acoustic resonators made out of a single AlN thin film employing a metal coverage of at least 50%, rivaling polycrystalline silicon capacitive resonators while exhibiting several orders of magnitude lower motional impedance [44], [45], [46].

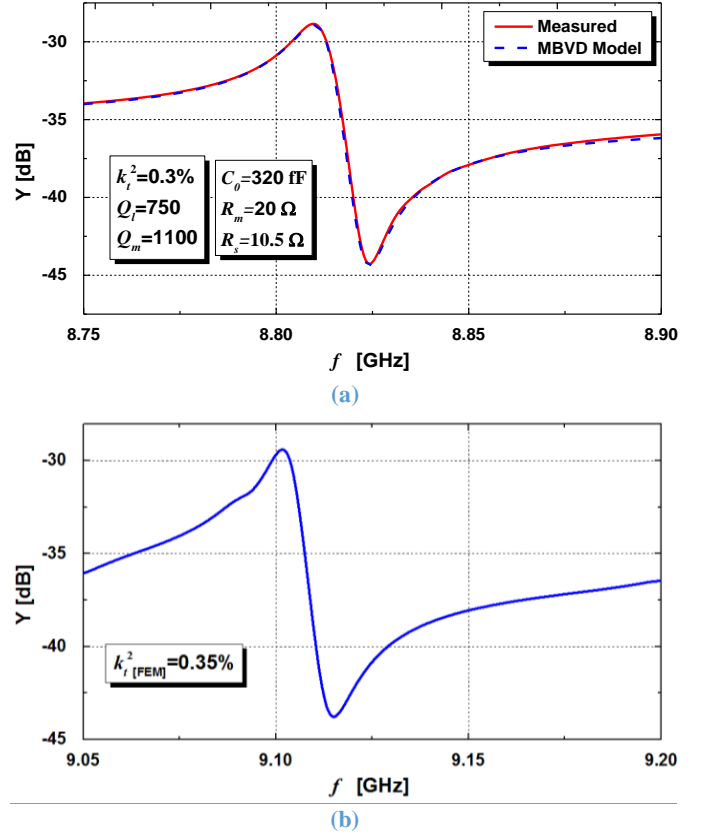

Fig. 8. (a) Measured and (b) FEM-simulated admittances of the sub-optimal COR design experimentally demonstrated: $t_{AlN}$=1 μm, $t_{Al}$=85 nm, $W$=1.15 μm, $\alpha$=57%, SW angle=60°.

A summary of $Q_l$-values reported in literature for piezoelectric resonators operating above 6 GHz is shown in Fig. 9. It is interesting to note that the all the demonstrated $Q_l$-values are nearly constant and below 400 in the 6-30 GHz range, even though AlN devices are supposed to show an increased $f$-$Q_{intrinsic}$ product in this frequency range [43]. This trend demonstrates how the increased loss associated with aggressive scaling of the piezoelectric film and the electrodes in conventional resonators poses a hard limit to the maximum achievable $Q_l$-values. The COR technology demonstrated in this work overcomes this fundamental limitation by exploiting the efficient transduction of a combined overtone mode of vibration in a relatively thick material stack, which directly translates into a groundbreaking improvement (>125%) in $Q$ factor compared to the state-of-the-art piezoelectric resonator technologies.

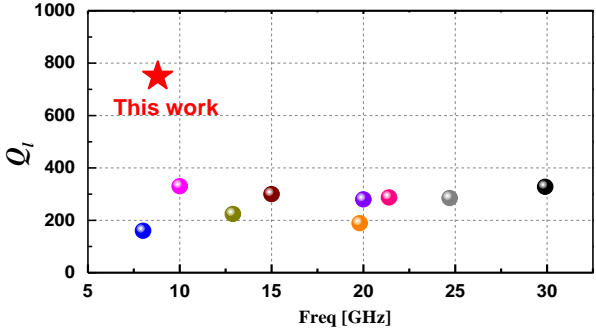

Fig. 9. A summary of $Q$-values for piezoelectric resonator technologies past 6 GHz [2], [8], [34], [47], [48],[49].

An electromechanical-coupling coefficient, $k_t^2 \sim 0.3\%$ was also extracted from the measured electrical response of the devices. It is worth noting that this experimental value is lower than the highest one predicted by FEM simulation for the optimal design ($t_{AlN}$=1 μm, $t_{Al}$=120 nm, $W$=1.15 μm, $\alpha$=35%, SW angle =90°, $k_t^2 \sim 0.78\%$). This is due to the fact that only a sub-optimal design ($t_{AlN}$=1 μm, $t_{Al}$=85 nm, $W$=1.15 μm, $\alpha$=57%, SW angle=60°, $k_t^2 \sim 0.35\%$) could be experimentally demonstrated because of the limited fabrication capabilities available. State-of-the-art lithographic techniques in Integrated Circuit (IC) industry would be sufficient to implement IDTs with pitches ($W$) and metallic coverage ($\alpha$) suitable for optimal operation in the mm-Wave range (see Tables I and II). Furthermore, the use of a SiO$_2$ hard mask for the AlN etch would be sufficient to improve the SW angle to >80° [27]. Nevertheless, the FEM simulation of the sub-optimal design (actual fabricated dimensions) matches well with the experimental results (FEM-simulated $k_t^2 \sim 0.35\%$), proving the validity of FEM model. It is also worth noting that demonstrated COR prototypes were properly sized to have a static capacitance, $C_0 \sim 320$ fF which corresponds to a termination impedance of 56 Ω at the device operating frequency, making the devices suitable for direct interface with 50 Ω radio frequency systems without using off-chip impedance matching networks.

## IV. CONTIGUOUS COR FILTERS FOR 5G mmWAVE SPECTRUM

Frequency scaling of existing micro-resonator technologies up to the mmWave range has not been successful to date due to the severe performance degradation associated with the required aggressive scaling in both the vertical and lateral dimensions of the resonator. This fundamental challenge, that has so far prevented the implementation of performing mmWave micro-acoustic filters, is overcome by the COR technology. In fact, thanks to the use of a combined overtone mode of vibration the operating frequency of CORs can be scaled up to mmWave frequencies while maintaining relaxed lithographic requirements (minimum feature >100 nm) and relatively thick AlN films (>220 nm) (Table I and Table II).

The FEM simulated operating frequencies of a COR and a BAW resonator, for different thicknesses of the AlN film, $t_{AlN}$, are shown in Fig. 10. The operating frequency of the COR is 73% higher than the one of the BAW resonators. As evident, the COR technology can deliver resonators operating up to 49

GHz while maintaining an AlN thickness > 200 nm (i.e. preserving good film quality of the piezoelectric film with <10% degradation compared to thicker films used in lover frequency devices) while the frequency of more conventional BAW resonators would be limited to 28 GHz under the same film thickness constraint.

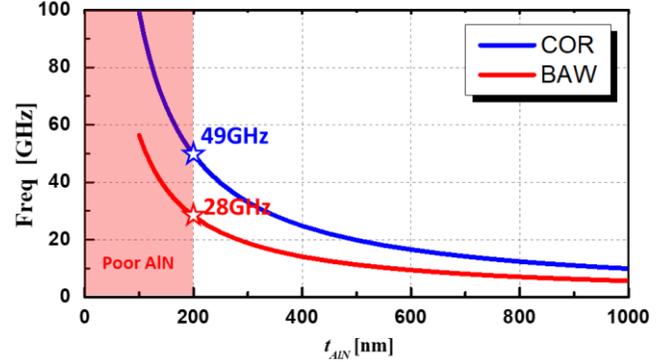

Fig. 10. Operating frequencies versus AlN thickness for CORs and BAW resonators. The metal electrodes were not included in the FEM models used in the simulations.

Another challenge associated with aggressive scaling of the film thickness is the highly increased sensitivity of the device operating frequency to process-related film thickness variations. This challenge is partially mitigated in CORs thanks to the transduction of a 2D mode of vibration with both lateral and vertical displacement components. As evident from the FEM simulation results in Fig 11, CORs show a lower frequency sensitivity to thickness variation compared to more conventional BAW devices.

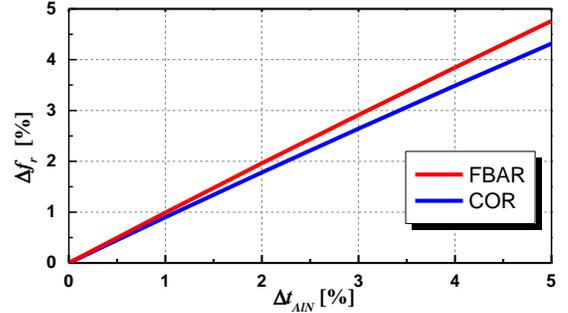

Fig. 11. Simulated relative frequency shift $\Delta f_r$ for different relative variations of the resonator thickness, $\Delta t_{AlN}$. Electrodes are not included in the simulations for a fair comparison.

The special 2-D modal characteristics of CORs also enable lithographic control of the device resonance frequency which is not possible in more conventional BAW resonators relying on a thickness extensional mode of vibration. In particular, our FEM simulations show that a lithographic frequency tuning of ~4%, with <20% reduction in $k_t^2$, is possible for both TFE-2 and LFE CORs operating at ~24 GHz (Fig. 12-13).

### TABLE I
#### DIMENSIONS FOR LFE CORs OF DIFFERENT FREQUENCIES

| Freq [GHz] | $W$ [nm] | $t_{AlN}$ [nm] | $t_{Al}$ [nm] |
|---|---|---|---|
| 24 | 431 | 375 | 45 |
| 25 | 414 | 360 | 43 |
| 26 | 398 | 346 | 42 |
| 27 | 383 | 333 | 40 |
| 28 | 370 | 321 | 39 |
| 29 | 357 | 310 | 37 |
| 30 | 345 | 300 | 36 |
| 31 | 334 | 290 | 35 |
| 32 | 323 | 281 | 34 |
| 33 | 314 | 273 | 33 |
| 34 | 304 | 265 | 32 |
| 35 | 296 | 257 | 31 |
| 36 | 288 | 250 | 30 |
| 37 | 280 | 243 | 29 |
| 38 | 272 | 237 | 28 |
| 39 | 265 | 231 | 28 |
| 40 | 259 | 225 | 27 |

### TABLE II
#### DIMENSIONS FOR TFE-2 CORs OF DIFFERENT FREQUENCIES

| Freq [GHz] | $W$ [nm] | $t_{AlN}$ [nm] | $t_{Al}$ [nm] |
|---|---|---|---|
| 24 | 422 | 383 | 42 |
| 25 | 405 | 368 | 40 |
| 26 | 389 | 354 | 39 |
| 27 | 375 | 341 | 37 |
| 28 | 361 | 329 | 36 |
| 29 | 349 | 317 | 35 |
| 30 | 337 | 307 | 34 |
| 31 | 326 | 297 | 33 |
| 32 | 316 | 288 | 32 |
| 33 | 307 | 279 | 31 |
| 34 | 298 | 271 | 30 |
| 35 | 289 | 263 | 29 |
| 36 | 281 | 256 | 28 |
| 37 | 274 | 249 | 27 |
| 38 | 266 | 242 | 27 |
| 39 | 259 | 236 | 26 |
| 40 | 253 | 230 | 25 |

### TABLE III
#### BANDWIDTH (BW) ACHIEVABLE BY 3RD ORDER TFE-2 COR LADDER FILTERS IN THE MMWAVE SPECTRUM

| Freq [GHz] | BW [MHz] |
|---|---|
| 24 | 355 |
| 25 | 370 |
| 26 | 385 |
| 27 | 399 |
| 28 | 414 |
| 29 | 429 |
| 30 | 444 |
| 31 | 459 |
| 32 | 473 |
| 33 | 488 |
| 34 | 503 |
| 35 | 518 |
| 36 | 533 |
| 37 | 547 |
| 38 | 562 |
| 39 | 577 |
| 40 | 592 |

### TABLE IV
#### BANDWIDTH (BW) ACHIEVABLE BY 3RD ORDER LFE COR LADDER FILTERS IN THE MMWAVE SPECTRUM

| Freq [GHz] | BW [MHz] |
|---|---|
| 24 | 150 |
| 25 | 156 |
| 26 | 163 |
| 27 | 169 |
| 28 | 175 |
| 29 | 181 |
| 30 | 188 |
| 31 | 194 |
| 32 | 200 |
| 33 | 206 |
| 34 | 213 |
| 35 | 219 |
| 36 | 225 |
| 37 | 231 |
| 38 | 238 |
| 39 | 244 |
| 40 | 250 |

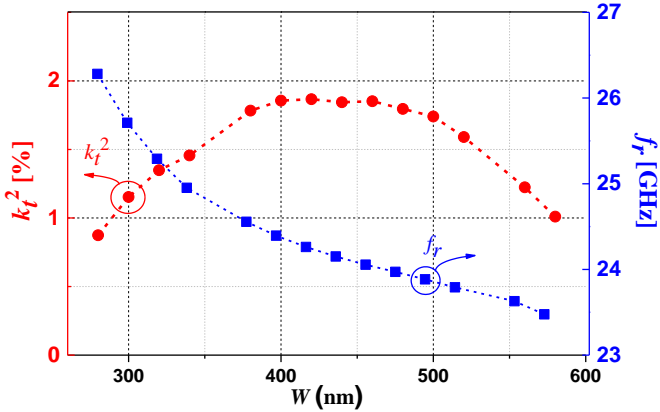

Fig. 12. Simulated resonant frequency and $k_t^2$ -values for different $W$-values for CORs using TFE-2 configuration around 24 GHz. The data points on $W$=360 nm and 540 nm are missing due to spurious modes.

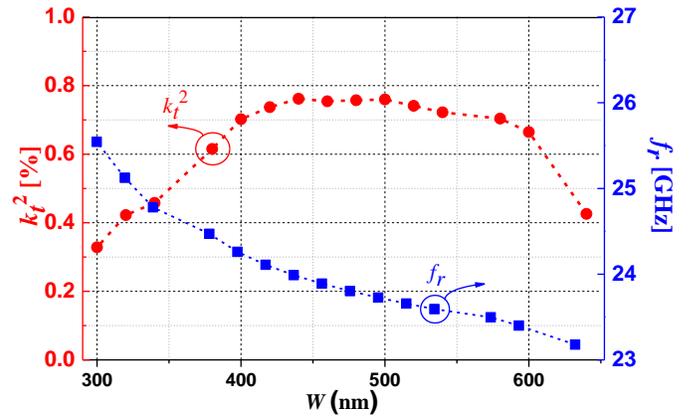

Fig. 13. Simulated resonant frequency and $k_t^2$ -values for different $W$-values for CORs using LFE configuration around 24 GHz. The data points on $W$=360 nm and 560 nm are missing due to spurious modes.

The capability of lithographically tune the operating frequency of CORs offers significant advantages for the implementation of filter architectures. In fact, the synthesis of a micro-acoustic filter requires the coupling (electrically or mechanically) of multiple resonators operating at different frequencies. The COR technology is capable of delivering such multi-frequency resonators on the same chip with reduced fabrication complexity compared to more conventional resonator technologies for which lithographic frequency tuning is not possible requiring additional process steps (i.e. additional lithography masks and selective material deposition) to adjust the frequencies of individual resonators used to synthesize the filter [50]. It is well-known that the bandwidth (BW) of a micro-acoustic ladder filter is directly proportional to the $k_t^2$-values of the resonators employed to synthesize the filter, while the filter Insertion Loss (I.L.) is inversely proportional to the product of the resonators $k_t^2$ and $Q$, as show in (12-13) [50].

$$BW \propto \frac{3}{\pi^2} k_t^2 \qquad (12)$$

$$I.L. \propto \frac{1}{k_t^2 Q} \qquad (13)$$

As discussed in the previous sections, AlN CORs provide $k_t^2$-values as high as 0.8%, for the LFE configuration, and 1.9% for the TFE-2 configuration, which directly translate into the capability of synthesizing mmWave filter with 100s MHz bandwidth, as reported in Table III and Table IV.

The performance level achievable by mmWave COR filters is here investigated by FEM simulations of contiguous 3rd order ladder filters synthesized by lithographically defined TFE-2 (Fig. 14) and LFE (Fig. 15) CORs operating at ~24 GHz. The simulation results show that 5 lithographically defined multi-frequency TFE-2 COR filters achieve an individual filter BW > 300 MHz capable of supporting an aggregated BW > 1.6 GHz. Similarly, 7 lithographically defined multi-frequency LFE COR filters achieve an individual filter BW > 150 MHz capable of supporting an aggregated BW > 1 GHz. It is also worth noting that, assuming a resonator $Q_l$ of 750, an I.L. < 1.2 dB is achieved with the TFE-2 configuration and an I.L. < 2.0 is achieved with the LFE configuration. Even with a degraded Q of 500 an IL<1.5 dB is achieved for the TFE-2 configuration and an IL<2.5 is achieved with the LFE configuration, as shown in Fig. 16.

It is worth noting, that this simulated performance level of the mmWave AlN LFE CORs is perfectly suitable for the synthesis of miniaturized and low-cost (simple 2-mask fabrication process) filters meeting the frequency, loss, bandwidth, roll-off and out-of-band rejection requirements of the 24 GHz 5G bands that have been recently licensed [51].

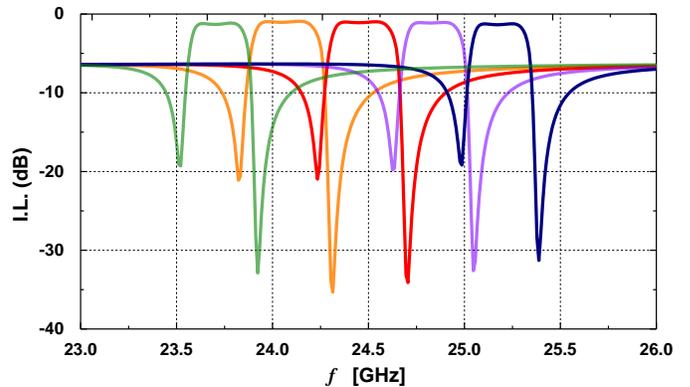

Fig. 14. FEM simulated performance of 5 contiguous ladder filters synthesized using lithographically defined multi-frequency TFE-2 CORs operating at ~24 GHz. Each filter has a BW> 300 MHz covering an aggregated BW> 1.6 GHz.

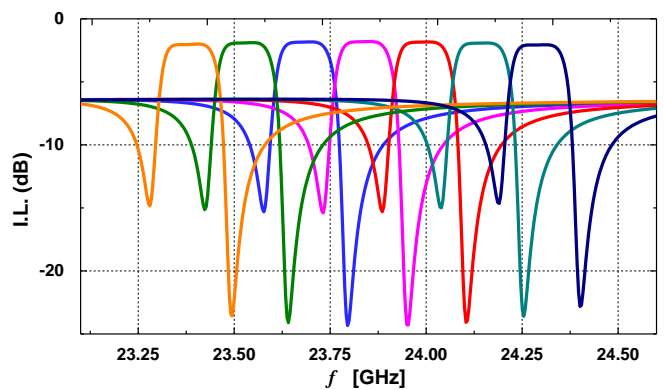

Fig. 15. FEM simulated performance of 7 contiguous ladder filters synthesized using lithographically defined multi-frequency LFE CORs operating at ~24 GHz. Each filter has a BW> 150 MHz covering an aggregated BW> 1 GHz.

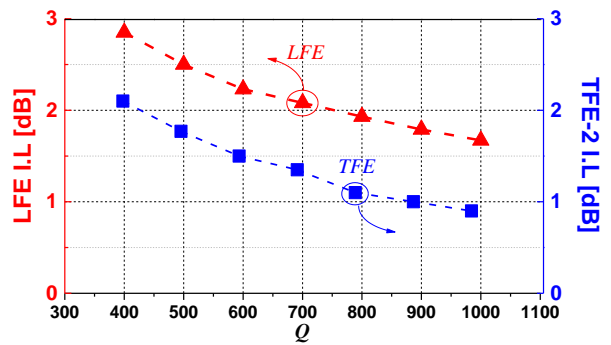

Fig. 16. Simulated Insertion Loss (I.L.) for 3rd order ladder filers made by LFE (red) and TFE-2 (blue) CORs for different $Q$.

## V. CONCLUSIONS

This paper presents a new class of AlN MEMS resonators, called *Combined Overtone Resonators* (CORs), which rely on the coherent combination of the $e_{33}$ and the $e_{15}$ piezoelectric coefficients of AlN for the multimodal excitation of the 2nd and 3rd order asymmetrical Lamb-wave overtones (A2 and A3 respectively) resulting into the efficient transduction of a 2-dimensional (2D) combined overtone mode (COM) of vibration in the cross-section of a suspended thin-film AlN plate. A COR prototype operating at ~9 GHz is fabricated and tested showing a high $k_t^2$~0.3% (using a simple top-electrode-only configuration with a 2-mask process) and a groundbreaking $Q$~1100 which is the highest ever achieved among piezoelectric resonators operating above 6 GHz. Detailed analytical and finite-element method (FEM) models are developed and used to demonstrate that a high electromechanical coupling coefficient, $k_t^2$~1.9%, can be achieved by optimizing the device geometry and electrode configuration. The unique advantages and great potential of the COR technology for the implementation of a new class of mmWave micro-acoustic filters is demonstrated by FEM analysis. 3rd order ladder filters, using lithographically defined CORs operating at ~24 GHz, are designed and their performance is simulated by FEM demonstrating the capability of synthesizing mmWave contiguous filters with 100s MHz bandwidths supporting aggregated bandwidths > 1 GHz. The demonstrated performance level of the mmWave AlN CORs is perfectly suitable for the synthesis of miniaturized, single-chip and low-cost contiguous filters meeting the frequency, loss, bandwidth, roll-off and out-of-band rejection requirements of the recently licensed 24 GHz 5G bands. Therefore, the proposed COR technology addresses the most important challenges that have so far prevented the implementation of high performance mmWave micro-acoustic filters, opening up a realm of new possibilities for a truly ubiquitous and pervasive mobile internet connectivity enabled by highly miniaturization and integrated 5G systems.

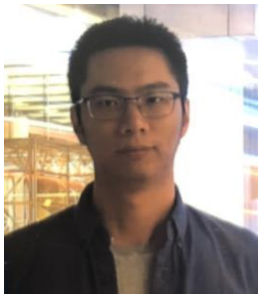

**Guofeng Chen** (S'16) received the Ph.D. degree in electrical and systems engineering from the Northeastern University, Boston, MA, USA in 2019. He had been a Research Assistant with the Northeastern Sensors and Nano Systems Laboratory, Boston, MA, USA, since 2015. He is currently an Electrical Staff Engineer with Skyworks Solutions, Inc., San Jose, CA, USA. His research interests include piezoelectric transducers such as micro-machined radio frequency resonators, filters, and ultrasonic transducers. He has co-authored 20 publications in those research areas.

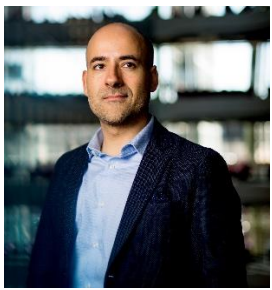

**Matteo Rinaldi** (S'08–M'10–SM'17) is an Associate Professor in the Electrical and Computer Engineering department at Northeastern University and the Director of *Northeastern SMART* a university research center that, by fostering partnership between university, industry and government stakeholders, aims to conceive and pilot disruptive technological innovation in devices and systems capable of addressing fundamental technology gaps in several fields including the Internet of Things (IoT), 5G, Quantum Engineering, Digital Agriculture, Robotics and Healthcare. Dr. Rinaldi received his Ph.D. degree in Electrical and Systems Engineering from the University of Pennsylvania in December 2010. He worked as a Postdoctoral Researcher at the University of Pennsylvania in 2011 and he joined the Electrical and Computer Engineering department at Northeastern University as an Assistant Professor in January 2012. Dr. Rinaldi's group has been actively working on experimental research topics and practical applications to ultra-low power MEMS/NEMS sensors (infrared, magnetic, chemical and biological), plasmonic micro and nano electromechanical systems, medical micro systems and implantable micro devices for intra-body networks, reconfigurable radio frequency devices and systems, phase change material switches, 2D material enabled micro and nano mechanical devices.

The research in Dr. Rinaldi's group is supported by several Federal grants (including DARPA, ARPA-E, NSF, DHS), the Bill and Melinda Gates Foundation and the Keck Foundation with funding of $14+M since 2012. Dr. Rinaldi has co-authored more than 130 publications in the aforementioned research areas and also holds 8 patents and more than 10 device patent applications in the field of MEMS/NEMS.

Dr. Rinaldi was the recipient of the IEEE Sensors Council Early Career Award in 2015, the NSF CAREER Award in 2014 and the DARPA Young Faculty Award class of 2012. He received the Best Student Paper Award at the 2009, 2011, 2015 (with his student) and 2017 (with his student) IEEE International Frequency Control Symposiums; the Outstanding Paper Award at the 18th International Conference on Solid-State Sensors, Actuators and Microsystems, Transducers 2015 (with his student) and the Outstanding Paper Award at the 32nd IEEE International Conference on Micro Electro Mechanical Systems, MEMS 2019 (with his student).